\documentclass[letter,11pt]{article}%

\usepackage{booktabs}
\usepackage{array, caption, threeparttable}
\usepackage[font=normalsize,labelfont=bf,labelsep=none]{caption}

\usepackage{indentfirst}
\usepackage[latin9]{inputenc}
\usepackage{geometry}
\usepackage{color}
\usepackage{amsmath}
\usepackage{amsthm}
\usepackage{amssymb}
\usepackage{graphicx}
\usepackage{caption}
\usepackage{setspace}
\usepackage{esint}
\usepackage{amsfonts}
\usepackage{comment}
\usepackage{booktabs}
\usepackage{lscape}
\usepackage{array}
\usepackage[titletoc,title]{appendix}
\usepackage[labelsep=period]{caption}
\usepackage{epstopdf}
\usepackage{natbib}
\usepackage[colorlinks,citecolor=blue,urlcolor=blue,filecolor=blue]{hyperref}
\usepackage{diagbox}
\usepackage{mathtools}
\usepackage{threeparttable}
\usepackage{multirow}
\usepackage{float}
\usepackage{bm}
\usepackage{siunitx}
\providecommand{\U}[1]{\protect\rule{.1in}{.1in}}

\providecommand{\U}[1]{\protect\rule{.1in}{.1in}}
\providecommand{\U}[1]{\protect\rule{.1in}{.1in}}

\numberwithin{equation}{section}
\makeatletter
\theoremstyle{plain}

\theoremstyle{definition}

\theoremstyle{plain}

\RequirePackage[colorlinks,citecolor=blue,urlcolor=blue]{hyperref}
\RequirePackage{hypernat}
\providecommand{\U}[1]{\protect\rule{.1in}{.1in}}
\providecommand{\definitionname}{Definition}
\providecommand{\propositionname}{Proposition}
\providecommand{\theoremname}{Theorem}
\providecommand{\definitionname}{Definition}
\providecommand{\propositionname}{Proposition}
\providecommand{\theoremname}{Theorem}
\providecommand{\definitionname}{Definition}
\providecommand{\propositionname}{Proposition}
\providecommand{\theoremname}{Theorem}
\makeatother
\providecommand{\definitionname}{Definition}
\providecommand{\propositionname}{Proposition}
\providecommand{\theoremname}{Theorem}

\graphicspath{{figures/}}
\bibpunct[, ]{(}{)}{,}{a}{}{,}

\usepackage{epsfig}

\usepackage{amssymb}

\usepackage{lineno}

\usepackage{booktabs}
\usepackage[font=normalsize,labelfont=bf,labelsep=period]{caption}
\usepackage{subcaption}
\usepackage{indentfirst}
\usepackage{array}
\usepackage{esint}
\usepackage{amsfonts}
\usepackage{comment}
\usepackage{booktabs}
\usepackage{lscape}
\usepackage{changepage}
\usepackage{array}
\usepackage{epstopdf}
\usepackage{natbib}
\usepackage{mathtools}
\usepackage{multirow}
\usepackage{bm}
\usepackage{siunitx}
\usepackage{amsmath}
\usepackage{amsthm}
\usepackage{appendix}

\usepackage[colorlinks,citecolor=blue,urlcolor=blue,filecolor=blue]{hyperref}

\usepackage{setspace} 
\usepackage{fullpage} 

\newcommand{\addtabletext}[1]{\raggedright #1}

\usepackage{ragged2e}

\begin{document}
\captionsetup[figure]{labelfont={bf},labelformat={default},labelsep=period,name={Fig.}}

\title{How does node centrality in a financial network affect asset price prediction? \thanks{The authors are listed in alphabetical order; and all authors made equal contributions.}}
\author{Yuhong Xu\thanks{Center for Financial Engineering and Math Center for
Interdiscipline Research, School of Mathematical Sciences, Soochow University,
P. R. China. This work is supported by the Natural Science
Foundation of China (No.12271391; No.11871050) and the Tang Scholar Fund.}, \ \ \ Xinyao Zhao\thanks{Corresponding author. Center for Financial Engineering, School of Mathematical Sciences, Soochow University, P. R. China. Email: xyzhao1996@stu.suda.edu.cn.}}

 \date{October 4, 2023}
 
\maketitle

\textbf{Abstract}: 
In complex financial networks, systemically important nodes usually play crucial roles. Asset price forecasting is important for describing the evolution of a financial network. Naturally, the question arises as to whether node centrality impacts the effectiveness of price forecasting. To explore this, we examine networks composed of major global assets and investigate how node centrality affects price forecasting using a hybrid random forest algorithm. Our findings reveal two counterintuitive phenomena: (i) factors with low centrality usually have better forecasting ability, and (ii) nodes with low centrality can be predicted more accurately in direction. These unexpected observations can be explained from the perspective of information theory. Moreover, our research suggests a criterion for factor selection: when predicting an asset price in a complex system, factors with low centrality should be selected rather than only factors with high centrality. Finally, we verify the robustness of our results using an alternative deep learning method.


\section{Introduction}
\label{Introduction}

Most recently, four US banks (Silver Gate Bank, Silicon Valley Bank,  Signature Bank, and First Republic Bank) collapsed and Credit Suisse was acquired. Unlike the four US banks, Credit Suisse is listed as one of the global systemically important banks by the Financial Stability Board.
In recent years, many scholars have employed complex networks and their centrality measures to identify the institutions \citep{Kuzubas, ChenHouJiang, QuLiuTang}.
Applications of complex networks in finance include correlation analysis of financial markets \citep{WangXieStanley, HanFanLing}, risk management \citep{Borochin and Rush, Ladley and Rousseau}, asset allocation \citep{Pozzi, Peralta, Vyrost, LiJiang, Olmo}, trading behavior analysis \citep{AdamicBrunetti, MusciottoPiilo}, among many others.
Systemically important financial institutions in the market typically have significant impacts on other institutions as well as the financial system as a whole. Particularly, if systemically important institutions are exposed to risk, the risk will easily spread to others. Hence, effectively and accurately predicting their prices contributes to regulating and controlling risks.

Naturally, we wonder whether there is a relationship between forecasting effectiveness and the topological structure of a complex network, especially node centrality.
Suppose that we enclose a collection of assets in a complex network and consider two problems.
(i) Are high central nodes more favorable than low central ones for more precise prediction results?
(ii) Are \textit{systemically important nodes} (SINs), i.e. nodes with high centrality, more predictable than systemically unimportant nodes?
Intuitively, nodes with high centrality are likely to contain more information about the entire system and better able to forecast asset prices.
Meanwhile, strong correlations between SINs and other nodes suggest possibly more reliable prediction results.

Surprisingly, our work shows that this is generally not the case. For problem (i), we divide the nodes into two sets with the highest/lowest centrality, respectively. When we forecast the price of the London gold spot, we find that the low-centrality set has a better forecasting ability. For problem (ii), the conclusion is that the \textit{least systemically important nodes} (LSINs) can be predicted more accurately than the \textit{most systemically important nodes} (MSINs). We explain these counterintuitive findings from the perspective of information theory. The set of factors with low centrality is shown to have more useful information and less noise for the forecast asset price. While predicting a SIN, we show that other nodes within the network have a lower information rate concerning the forecast asset.

In recent years, some scholars have investigated the relationship between node centrality and asset allocation. \cite{Pozzi} were the first to apply the topological properties of financial networks to portfolio selection. Their empirical findings suggested that a higher allocation of stocks with lower centrality within the network can mitigate risks. Subsequently, \cite{Peralta} established a theoretical foundation for the inverse correlation between the wealth allocated to a specific asset and its eigenvector centrality. \cite{Vyrost} further enhanced portfolio optimization by introducing a constraint where asset weights must align with the inverse of asset centrality within a given network. Their study, employing various network structures and centrality metrics, demonstrated substantial improvements in risk-minimized portfolios.

Parallel research efforts have been made by \cite{LiJiang} and \cite{Olmo}. Moreover, some scholars have leveraged complex networks to forecast asset prices, as evidenced by works such as \cite{WangTian2016}, \cite{WangZhao2018}, \cite{WangZhang2019}, \cite{XuWang2020}, \cite{Hu1,Hu2}, and \cite{Ribeiro}. However, to the best of our knowledge, our work is the first attempt to investigate the relationship between node centrality and effectiveness of asset price prediction.

In this paper, we select 37 major global assets including London gold spot and its related assets. We construct complex networks based on the correlations between their prices. For problem (i), we take the London gold spot as a predicted target. We use the first and last several central nodes as the factors to explore the effects of the factors' centrality on the prediction. We employ a hybrid random forest algorithm based on time-varying filtering-based empirical mode decomposition (TVF-EMD-RF) to predict asset prices. By comparing the forecasting results of the three sets of the first/last 6, 8, and 10 central nodes as factors, we find that all level accuracies using the last several central nodes are better than those using the first several central nodes. In particular, directional accuracies using the last several nodes as factors are, on average, over 4\% higher than those using the first central nodes. Therefore, we conclude that factors with low centrality generally have better forecasting ability than those with high centrality. To provide a reasonable interpretation for the counterintuitive phenomenon, further empirical research employing information theory reveals that factors with low centrality have more useful information and less noise for the forecast asset price. For problem (ii), we select the assets with the most occurrences in the first and last 5 central nodes, respectively, as the forecast nodes to analyze the impact of the forecast nodes' centrality on the prediction. When we forecast an asset price, we extract the top 10 principal components of the remaining 36 nodes as the factors. With the help of TVF-EMD-RF, we find that the LSINs can be forecast more accurately than the MSINs in terms of direction. The directional accuracies of the former are approximately 5\% higher than those of the latter. Using information theory again, we find that the former have higher information rates. This may provide a reasonable explanation for the finding. Finally, we use an alternative deep learning algorithm, TVF-EMD-DELM, to verify the robustness of our main conclusions.

The remainder of this paper is organized as follows. Section \ref{Sec 2} introduces the main methods and concepts, including the complex network and eigenvector centrality, forecasting approaches, forecasting steps, and evaluation criteria.  Section \ref{Sec 3}  explores the impact of the factors' centrality on the prediction. Section \ref{Sec 4} explores the effects of the centrality of the forecast nodes on the prediction. Section \ref{Sec 5} examines the robustness of our findings. Finally, Section \ref{Sec 6} concludes the paper.

\section{Concepts and Methods}
  \label{Sec 2}

\subsection{Complex network and eigenvector centrality}
Complex networks are graphs consisting of nodes and edges denoted as $G=(N, E)$. \(G\) is undirected if all its edges do not have direction and weighted if all its edges are weighted.

	To establish a complex network, an \textit{adjacency matrix} $A=(a_{ij})_{n\times n}$ should be first constructed based on a particular relationship among nodes, where $n$ is the number of nodes and $a_{ij}$ represents the connection strength between nodes $i$ and $j$. In this paper, we construct dynamic complex networks of major global assets based on the Pearson correlation coefficients between two asset prices using sliding time windows.
    At time $t$, we calculate correlation coefficients from $t-m+1$ to $t$, where $m$ is the window size. As is well known, for two series $u=\langle u_k\rangle$ and $v=\langle v_k\rangle$, their correlation coefficients are obtained in the following way \citep{Feller},
    \vspace{-2mm}
	\begin{equation}	
\begin{aligned}
		\centering
		\rho_t(u,v):&=\frac{Cov_t(u,v)}{\sqrt{D_t(u)}\sqrt{D_t(v)}}\\ \notag
  &=\cfrac{\sum_{k=t-m+1}^{t}(u_k-\overline{u})(v_k-\bar{v})}{\sqrt{\sum_{k=t-m+1}^{t}(u_k-\bar{u})^2}\sqrt{\sum_{k=t-m+1}^{t}(v_k-\bar{v})^2}},
	 \end{aligned}
\end{equation}
	where $\bar{u}$ and $\bar{v}$ are the mean values of the series $u$ and $v$, respectively, in $m$ days. Then, the Pearson correlation coefficient matrix is defined as

\begin{equation}
   		Corr(t)=\left[
		\begin{array}{ccccc}
			\rho_t(X_1,X_1) &\rho_t(X_1,X_2) &\cdots &\rho_t(X_1,X_l)  &\rho_t(X_1,Y) \\
			\rho_t(X_2,X_1) & \rho_t(X_2,X_2)&\cdots &\rho_t(X_2,X_l)  &\rho_t(X_2,Y)\\
			\cdots  & \cdots &\cdots &\cdots &\cdots\\
			\rho_t(X_l,X_1) & \rho_t(X_l,X_2)&\cdots &\rho_t(X_l,X_l)  &\rho_t(X_l,Y)\\
			\rho_t(Y,X_1) & \rho_t(Y,X_2)&\cdots &\rho_t(Y,X_l)  &\rho_t(Y,Y)\\
		\end{array}
		\right],
 \label{corr}
\end{equation}
 where $X_1, X_2, \cdots, X_l$ represent factors and $Y$ represents the forecast series. The elements of the adjacency matrix are the absolute values of the corresponding elements of the matrix $Corr(t)$ because positive and negative correlations are considered equally important.

\begin{figure}[htbp]
		\centering
	\includegraphics[width=4.5 in]{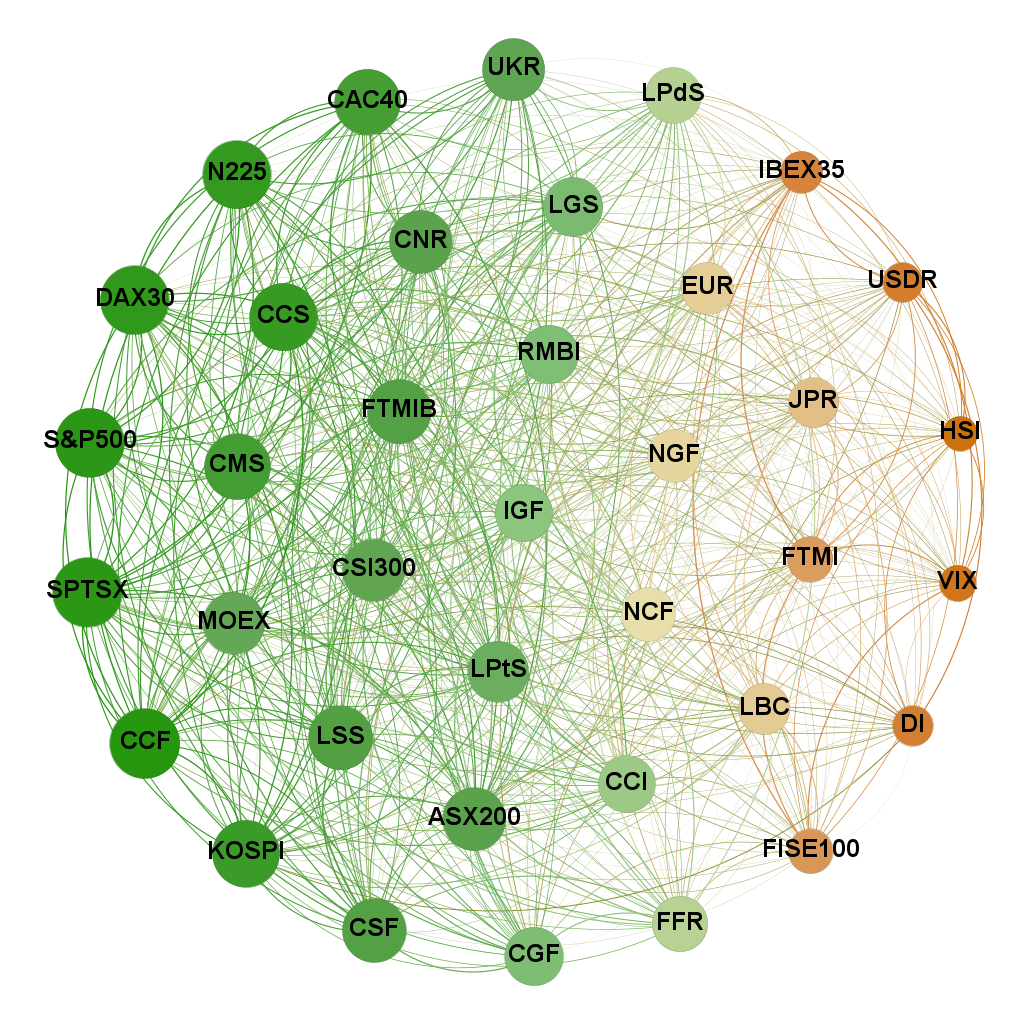}
		\centering
		\caption{Major global assets network. Nodes are labeled by abbreviations of the associated assets (Table \ref{data source} in  \ref{appendix:Data}). The bigger a node and the darker green it is, the higher its eigenvector centrality and the more important it is in the complex network. The darker orange it is, the weaker its centrality is. The heavier the line, the greater its weight and the stronger the correlation between the two nodes. Otherwise, the color of a line is a mixture of the colors of the two nodes that it connects. As we see, the London gold spot (LGS) has a moderate centrality.}
		\label{network}
	\end{figure}

   Appropriate utilization of topological properties greatly facilitates the analysis of complex networks. In particular, centrality measures characterize the centrality and importance of the nodes in a network. For undirected weighted networks, \textit{eigenvector centrality} is an applicable centrality measure \citep{Peralta, Olmo}. The basic idea of eigenvector centrality is that the centrality of a node is a function of the centrality of its adjacent nodes \citep{Bonacich}. It is defined as follows,
    	\begin{equation}
		EC_i=x_i=\frac{\sum_{j=1}^{n} a_{ij} x_j}{\lambda},
		\label{EC}
	\end{equation}
	where $\lambda$ is the largest eigenvalue of $A$, $x=(x_1,\ldots,x_n)^T$ is the eigenvector corresponding to $\lambda$, and \(x_i\) represents the eigenvector centrality of node $i$. This indicates that the importance of a node depends on the relationship between the node and its neighbors (nodes directly connected with it) and the importance of its neighbors. See Fig. \ref{network} for a complex network consisting of major global assets. The larger a node is and the darker green it appears, the higher its eigenvector centrality is within the complex network. Likewise, the thicker the line, the greater its weight and the stronger the correlation between the two nodes.

\subsection{Forecasting approaches}

\begin{figure}[ht]
		\centering
	\includegraphics[width=4.5 in]{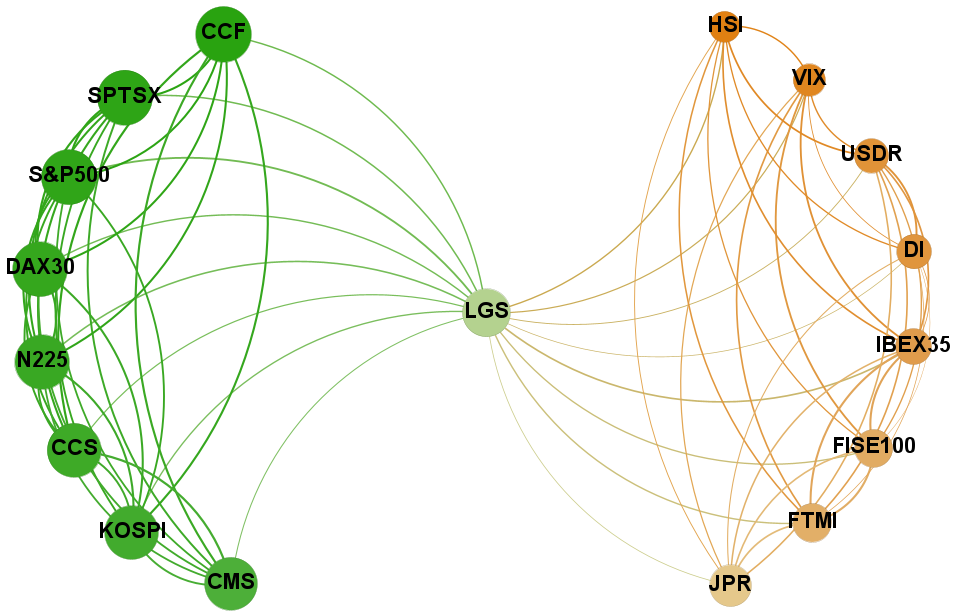}
		\centering
		\caption{Problem (i). The left 8 green nodes are the first 8 central nodes; the right 8 orange nodes are the last 8 central nodes. Nodes are labeled by abbreviations of the associated assets. We use them respectively to forecast the price of the London gold spot. }
		\label{Problem 1}
	\end{figure}

Since all the other nodes in the network are effective factors for forecasting the price of London gold spot,
for problem (i), we take the London gold spot as the forecast object and separately use the first $n$ and last $n$ central nodes as factors to explore the influence of the factors' centrality on the prediction results, which is illustrated in Fig. \ref{Problem 1}. We use TVF-EMD-RF to predict the gold price and compare forecasting results of the first/last several central nodes as factors. More details about the random forest algorithm and TVF-EMD method are described in \ref{appendix:Methods}.

For problem (ii), we select the assets that appear most frequently in the first and last 5 central nodes as the forecast nodes and analyze the impact of the forecast nodes' centrality on the prediction results. When forecasting an asset price, we use the remaining nodes as factors.

\subsection{Forecasting steps}
    To optimize the forecasting performance, we use machine learning algorithms with TVF-EMD in a rolling forecasting method. The rolling window is fixed for $m$ days and recursively changes as the forecasting progresses in time. The steps are as follows: (i) Select proper influential factors as features according to the correlation or other criteria (the factors can vary) and historical $a$-day data of the forecast asset price. (ii) Divide the dataset into training and testing sets. For the rolling forecasting model, the previous $m$-day data of the forecast day are set as the training set. (iii) For each window, we extract a fixed amount of intrinsic mode functions (IMFs) from the forecast time series using TVF-EMD. (iv) Forecast each IMF using machine learning algorithms and sum the forecast values as the final value.

\subsection{Evaluation criteria}	

Directional accuracy of asset price prediction is important for investment decision and risk management. The direction statistic is defined as
	
	\begin{equation}
	Dstat=\frac{\sum_{t=1}^{T} I_t}{T} \times 100\%,
		\label{DS}
	\end{equation}
	where

	\[
	I_t=\left\{
	\begin{array}{ll}
		1,  \quad if \quad (Y_{t}-Y_{t-1})(\hat{Y}_{t}-Y_{t-1}) \geq 0;\\
		0,  \quad if \quad (Y_{t}-Y_{t-1})(\hat{Y}_{t}-Y_{t-1})<0.
	\end{array}
	\right.
	\]
	 $T$ is the number of elements in the testing set, $Y_{t}$ is the true value of the testing sample, and $\hat{Y}_{t}$ is the forecast value at time $t$. The larger the value, the more accurate the forecasting results.
	
	To fully compare the forecasting results, we also use MAPE, MAE, RMSE, and R$^2$ to evaluate the prediction accuracy (see Table \ref{Loss functions}). A high R$^2$ value is preferable for the prediction. For the other three criteria, lower values indicate more accurate forecasting results.

	\begin{table}[ht]
		\centering
		\caption{Accuracy evaluation criteria.}
  \begin{tabular}{lll}
\toprule
  Evaluation criteria & Abbreviation   &Formula \\
			\hline
			\specialrule{0em}{2pt}{2pt}
			Mean absolute percentage error & MAPE & $\frac{1}{T} \sum_{t=1}^{T} |\frac{\hat{Y}_{t}-Y_{t}}{Y_{t}}|$\\
			\specialrule{0em}{2pt}{2pt}
			Mean absolute error &  MAE &$\frac{1}{T} \sum_{t=1}^{T}|\hat{Y}_{t}-Y_{t}|$ \\
			\specialrule{0em}{2pt}{2pt}
			Root mean square error & RMSE & $\sqrt{\frac{1}{T}\sum_{t=1}^{T} (\hat{Y}_{t}-Y_{t})^2}$ \\
\specialrule{0em}{2pt}{2pt}
    Goodness of fit & R$^2$ &  $ 1-\frac{\sum_{t=1}^{T}(\hat{Y}_{t}-Y_{t})^2}{\sum_{t=1}^{T}(\bar Y_{t}-Y_{t})^2}$ \\
			\bottomrule

  \end{tabular}
\label{Loss functions}
	\end{table}

   \section{The impact of factors' centrality on prediction}	

   \label{Sec 3}

   In this paper, we select 37 global assets, including 14 commodities, 14 stock indices, 6 currency rates, and 3 indicators (Dollar Index, RMB Index, VIX) to construct networks of major global assets. Table \ref{data source} in   \ref{appendix:Data} lists their names, categories, abbreviations, and data sources. Considering the training cost and model evaluation requirements, the training set lasts 1,200 days, and the rolling forecasting time of each testing set is set as $300$ days. Two testing sets are constructed with the first one covering data from March 5, 2019 to April 28, 2020 (dataset I), and the second from February 5, 2021 to March 31, 2022 (dataset II). To forecast the asset price at time $t$, we also use its historical 2-day prices.

   When constructing the network, we choose a sliding window of 1,200 days consistent with the length of training sets.
   Fig. \ref{network} shows the asset network using data from August 18, 2017 to March 30, 2022.  During this period, the first 10 central factors are the COMEX copper futures, SPTSX, S\&P500, DAX30, N225, CRB comprehensive spot, KOSPI, CRB metal spot, CAC40, and London silver spot. The last 10 central factors are the HSI, VIX, USD effective exchange rate, dollar index, IBEX35, FISE100, FTMI, JP effective exchange rate, London brent crude, and EU effective exchange rate.

In this section, we study the effects of factors' centrality on predicting outcomes with the London gold spot as the forecast node.

   \subsection{Forecasting results}

   To forecast the price at time $t$, we select features according to the eigenvector centrality of each node at time $t-1$ to avoid the information leak problem. Each node in the asset network has a different eigenvector centrality on a daily basis, which reflects the significance or status of each node during the preceding window.

  To discuss the impact of the factors' centrality on the forecasting results, we consider three special cases where we use the first 6, 8, and 10 central nodes and the last 6, 8, and 10 central nodes, except the London gold spot, as factors each day and compare their forecasting results on the testing set. Table \ref{I Results RF} reports the results for the two experimental datasets obtained by TVF-EMD-RF. However, the last 6, 8, and 10 central factors have markedly better forecasting performance in terms of both directional and level accuracy than the first 6, 8, and 10 central factors do, respectively.
  To be specific, directional accuracies using the last several central factors are generally 2\%-6\% higher than those using the first several factors especially 8 central factors, and MAPEs are lower at the same time. Taking the dataset I for instance, the directional accuracy when using the last 8 central factors is 61.33\%, which is 5.00\% higher than that when using the first 8 central factors. Hence, the factors' centrality has a significant effect on the prediction.

 \begin{table*}[htbp]
   \renewcommand\arraystretch{1.2}
		\centering
		\caption{Forecasting results of the London gold spot price obtained by TVF-EMD-RF.}
		\begin{tabular}{llcccc}
       \toprule
			Testing set  &Factors &Dstat (\%)&MAPE (\%) &MAE&RMSE \\
		\hline
			\multirow{6}{*}{Dataset I}
			&First 6 central factors & 55.67	&1.01	&15.18 		&21.22   \\
  			&\textbf{Last 6  central factors}   &58.33	&0.96  &14.40 &	19.82   \\
			\cline{2-6}
		    &First 8 central factors    & 56.33	&1.16	&17.65	&	25.87   \\
			&\textbf{Last 8  central factors}  &61.33 &1.05	&15.76 	&22.87   \\
			\cline{2-6}
			&First 10  central factors    &54.00	&1.11	&16.88 &23.24   \\
   			&\textbf{Last 10  central factors}  &56.00 &1.07 &16.22 &22.27 \\
			\hline \hline
			\multirow{6}{*}{Dataset II}
		&First 6  central factors  & 54.00	&0.86	&15.52 	&21.23  \\
			&\textbf{Last 6  central factors}   &59.00	&0.86  &15.56	 &	20.86   \\
			\cline{2-6}
		&First 8 central factors   & 56.00	&0.84	&15.32	&	20.25    \\
			&\textbf{Last 8  central factors}  &61.67 &0.83	&15.10	&23.75 \\
   			\cline{2-6}
			&First 10  central factors    & 56.33	&0.87	&15.72 &21.37   \\
			&\textbf{Last 10  central factors}  &58.33 &0.87	&15.74	&20.93  \\
   \bottomrule
		\end{tabular}
  		\centering
		\label{I Results RF}
	\end{table*}

To further substantiate our findings, we consider an additional 12 testing sets starting on 2017/08/18, each spanning a hundred days. We utilize the central factors from the first eight and the last eight factors to forecast the London gold spot price. As indicated in Table \ref{100day},  in 11 out of the testing sets, the predictive performance of the last 8 central factors in terms of direction is significantly better than that of the first 8 central factors, with the exception of the second testing set. Furthermore, whether considering a window of every two or three hundred days, the predictive ability of the last 8 central factors consistently outperforms that of the first 8 central factors.

\begin{table*}[htbp]
   \renewcommand\arraystretch{1.2}
		\centering
		\caption{Direction statistic (\%) of the London gold spot price prediction in 1200 days obtained by TVF-EMD-RF using the statistic window lengths $w=100, 200,$ and $300$, respectively.}
		\begin{tabular}{ccccccc}
       \toprule
			\multirow{2}{*}{Testing set}   &\multicolumn{3}{c}{\textbf{First 8 central factors}} &\multicolumn{3}{c}{\textbf{Last 8 central factors}} \\
   		\cline{2-7}
 &  $w=100$ &$w=200$ & $w=300$   &$w=100$ &$w=200$ & $w=300$\\
		\hline
       1     & 58.00 & \multirow{2}{*}{57.50$^*$} & \multirow{3}{*}{55.67$^*$$^*$} & 63.00 & \multirow{2}{*}{59.00$^*$} & \multirow{3}{*}{59.00$^*$$^*$} \\
    2     & 57.00 &       &       & 55.00 &       &  \\
    3     & 52.00 & \multirow{2}{*}{52.00$^*$} &       & 59.00 & \multirow{2}{*}{58.00$^*$} &  \\
    4     & 52.00 &       & \multirow{3}{*}{54.67$^*$$^*$} & 57.00 &       & \multirow{3}{*}{60.67$^*$$^*$} \\
    5     & 53.00 & \multirow{2}{*}{56.00$^*$} &       & 59.00 & \multirow{2}{*}{62.50$^*$} &  \\
    6     & 59.00 &       &       & 66.00 &       &  \\
    7     & 57.00 & \multirow{2}{*}{56.00$^*$} & \multirow{3}{*}{56.00$^*$$^*$} & 59.00 & \multirow{2}{*}{58.00$^*$} & \multirow{3}{*}{58.67$^*$$^*$} \\
    8     & 55.00 &       &       & 57.00 &       &  \\
    9     & 56.00 & \multirow{2}{*}{54.50$^*$} &       & 60.00 & \multirow{2}{*}{60.00$^*$} &  \\
    10    & 53.00 &       & \multirow{3}{*}{56.00$^*$$^*$} & 60.00 &       & \multirow{3}{*}{61.67$^*$$^*$} \\
    11    & 61.00 & \multirow{2}{*}{57.50$^*$} &       & 65.00 & \multirow{2}{*}{62.50$^*$} &  \\
    12    & 54.00 &       &       & 60.00 &       &  \\
     \bottomrule
		\end{tabular}

     \addtabletext{\small{Notes: $^*$ The statistic window length is $w=200$ days; $^*$$^*$ The statistic window length is $w=300$ days.}}
		\label{100day}
	\end{table*}

 Basically, when forecasting the London gold spot price, we can come to a conclusion: factors with low centrality are more forecasting-effective than those with high centrality. It is worth noting that the high centrality of nodes in a complex network is not equivalent to strong correlations between nodes and the forecast object. Thus our findings do not conflict with conventional correlation-based factor selection methods. Additionally, they suggest a fresh idea for factor selection which recommends factors with low centrality should be selected rather than just those with high centrality when predicting asset prices in a complex financial system. This is somewhat similar to the suggestion by \cite{Pozzi} who claim that more stocks with low centrality in the network should be allocated to avoid risks.

  In the remainder of this section, we aim to explain our main conclusions from an information perspective.
	
  \subsection{Principal component analysis for the selected factors}
	
  We first perform principal component analysis (PCA) for factors with high and low centrality.

\begin{figure}[htbp]
    \centering
    \begin{subfigure}{0.47\textwidth}
        \centering
        \includegraphics[width=\linewidth]{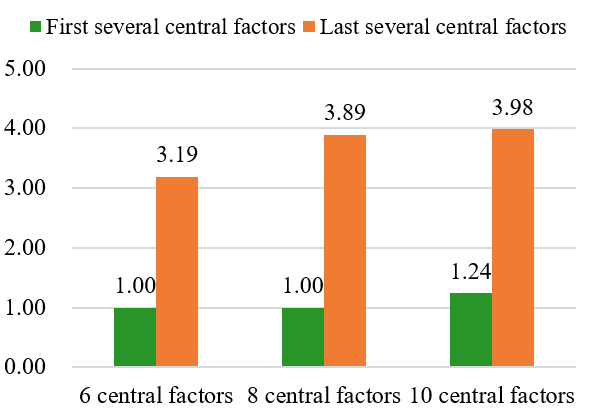}
  \subcaption{At 80\% contribution rates.}

    \end{subfigure}%
    \hfill
    \begin{subfigure}{0.47\textwidth}
        \centering
        \includegraphics[width=\linewidth]{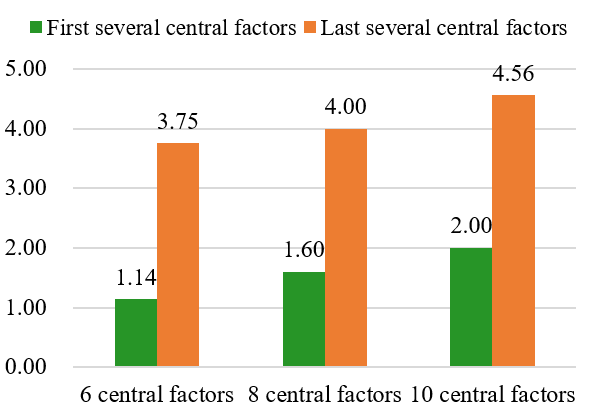}
    \subcaption{At 85\% contribution rates.}
    \end{subfigure}
\caption{Average quantities of principal components in the dataset I.}
\label{PCA quantity}
\end{figure}

\begin{figure}[htbp]
    \centering
    \begin{subfigure}{0.47\textwidth}
        \centering
        \includegraphics[width=\linewidth]{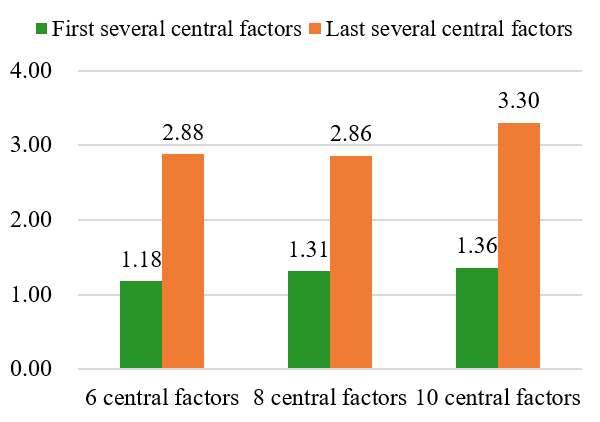}
  \subcaption{At 80\% contribution rates.}

    \end{subfigure}%
    \hfill
    \begin{subfigure}{0.47\textwidth}
        \centering
        \includegraphics[width=\linewidth]{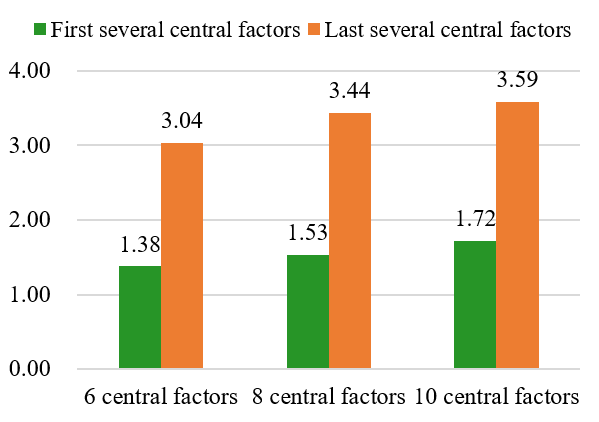}
    \subcaption{At 85\% contribution rates.}
    \end{subfigure}
\caption{Average quantities of principal components in the dataset II.}
\label{PCA quantity2}
\end{figure}

 For each testing sample, we extract the principal components of the first/last 6, 8, and 10 central factors from the preceding 1201 days at various contribution rates. In this way, we obtain a 300-day time series for every experimental dataset regarding the number of principal components for the first/last central factors. Fig. \ref{PCA quantity} shows the average quantities of principal components in the dataset I. The factors with low centrality always have more principal components than those with high centrality for the same quantity of factors at the same contribution rate. For example, the first 8 central factors have only one principal component at the 80\% contribution rate, whereas the last 8 central factors have 3.89 principal components. Similar results for the dataset II are shown in Fig. \ref{PCA quantity2}.

 This may indicate that the set of factors with high centrality is less informative. A possible reason is that factors with high centrality have strong internal correlations among themselves. Hence, they have fewer principal components and contain a small amount of information. To confirm the above results, we calculate the mutual information between factors and the forecast node in the next subsection.

	\subsection{Estimating entropy and mutual information of factors and the forecast node}
  As is well-known, the PCA method considers only the information that factors contain, focusing on the relationship among factors and ignoring the relationship between factors and the forecast asset. Selecting features with maximal relevance and minimal redundancy is a good feature selection approach, and features selected in this manner usually have good prediction or classification results \citep{Ding2005, Peng}. Relevance is typically characterized in terms of correlation or mutual information. Therefore, to further explore why the set of factors with low centrality can better predict gold prices, we analyze it from the point of view of information theory.

  Entropy reflects the expected amount of information of a random variable and measures its uncertainty. \textit{Differential entropy} is defined as

    \begin{equation}
		h(X):=-\int_\mathcal{X} p(x)\ln p(x)\,dx,
       \label{h(X)}
   \end{equation}
    where $X$ is a continuous random variable with a probability density function $p$ whose support is a set $\mathcal{X}$. \textit{Mutual information} assesses the mutual dependence between two random variables, while conditional entropy measures the level of remaining uncertainty of a random variable given that the other random variable has been learned. For more details, please refer to \ref{appendix:information}. Let $I(X,Y)$ and $h(X|Y)$ stand for the mutual information between random variables $X$ and $Y$, and \textit{conditional differential entropy} of $X$ given $Y$, respectively. Note that the mutual information is equal to the difference between differential entropy and conditional differential entropy.

    Let $\emph{\textbf{X}}$ denote the factors and $Y$ denote the gold price. We resort to the Gaussian kernel density method in the estimation of the differential entropy for $\emph{\textbf{X}}$ and $Y$, and mutual information between $\emph{\textbf{X}}$ and $Y$. See \cite{Ahmad} and \cite{Moon} for the detailed procedure.

    First, we estimate the differential entropy of the gold price under a rolling window of 1,201 days and obtain a 300-day time series for each experimental dataset. The average differential entropy is 5.86 in the dataset I, and it is 6.76 in the dataset II. The average differential entropy of the dataset II is larger, indicating that the uncertainty of the price is greater in the dataset II.

   Second, to measure the amount of information that a set of factors $X_1, X_2, \cdots, X_n$ provides to the forecast asset $Y$, we need to estimate the mutual information $\small{I(X_1, X_2, \cdots, X_n; Y)}$ instead of the sum of the mutual information between each factor $X_i$ and $Y$,  namely $\sum_{i=1}^n I(X_i, Y)$. This is because the mutual information between different pairs of factors may overlap according to Eq. \ref{MI2} in \ref{appendix:information}. Mutual information between continuous factors $X_1, X_2, \cdots, X_n$ and $Y$ supported at $\mathcal{X}_1, \mathcal{X}_2, \cdots, \mathcal{X}_n$ and $\mathcal{Y}$ is expressed as follows,
   \begin{align}
\label{I}
				I(\emph{\textbf{X}};Y)&= I(X_1, X_2, \cdots, X_n; Y)\nonumber\\&=h(X_1, X_2, \cdots, X_n)-h(X_1, X_2, \cdots, X_n|Y) \\
&=\int_{\mathcal{X}_1}\int_{\mathcal{X}_2}\cdots\int_{\mathcal{X}_n}\int_{\mathcal{Y}}p(x_1, x_2, \cdots, x_n, y)\ln\frac{p(x_1, x_2, \cdots, x_n, y)}{p(x_1, x_2, \cdots, x_n)p(y)}\,dx_1dx_2\cdots dx_ndy. \nonumber
\label{I}
\end{align}
where $p(x_1, x_2, \cdots, x_n)$ and $p(x_1, x_2, \cdots, x_n, y)$ are the joint probability density functions of $\emph{\textbf{X}}$ and ($\emph{\textbf{X}}$, $Y$), respectively. The quantity of information that these factors contain regarding the gold price increases with the mutual information $I(\emph{\textbf{X}};Y)$ between them,

Third, we should consider the proportion of valid to invalid information that $n$ factors bring to the forecast target. The larger the amount of invalid information, the greater the interference to forecast $Y$. Subtracting the mutual information $I(\emph{\textbf{X}};Y)$ from the differential entropy of $n$ factors gives a noise term, which is exactly $h(X|Y)$. Then the \textit{information/noise ratio} of $\emph{\textbf{X}}$ to $Y$ can be defined as
	\begin{equation}
		 I/N=\frac{I(\emph{\textbf{X}}; Y)}{h(\emph{\textbf{X}}|Y)}\times 100\% =\frac{I(X_1, X_2, \cdots, X_n; Y)}{h(X_1, X_2, \cdots, X_n|Y)}\times 100\%.
        \label{I/N}
	\end{equation}
    Evidently, $I/N$ is positively related to $I(\emph{\textbf{X}};Y)$ and negatively related to the conditional differential entropy of $n$ factors.

    An ideal situation is that the information that the factors provide to the forecast asset price $Y$ exactly equals the information of $Y$ (the entropy of $Y$). That is, the factors of $Y$ have no interference. In this case, the directional accuracy will be 100\%, and the error will be 0.

	\begin{table}[ht]
      \renewcommand\arraystretch{1.2}
		\centering
		\caption{Differential entropy, mutual information, conditional differential entropy, and information/noise ratio, all in average sense.}
		\begin{tabular}{lllllll}
  \toprule
			\multirow{2}{*}{Testing set}&\multirow{2}{*}{Factors}   &\multicolumn{4}{c}{Daily average} \\
	\cline{3-6}
 & &   $h(\emph{\textbf{X}})$	&	$I(\emph{\textbf{X}};Y)$&$h(\emph{\textbf{X}}|Y)$ & $I/N$ \\
	    \hline 		
			\multirow{2}{*}{Dataset I}
                &First 8 central factors   &48.96 &1.42 &47.54 &2.99\%	  \\
			&Last 8 central factors    &14.48 &1.49 & 12.98 &11.67\%\\ 
		\hline \hline
			\multirow{2}{*}{Dataset II}
   	      &First 8 central factors 	&43.74 &1.47 & 42.27  &3.48\%   \\
			&Last 8 central factors   &27.22 &1.52 &25.72 	&6.70\% \\ 
				\bottomrule
		\end{tabular}
		\label{MI IN}

\vspace{2mm}
\raggedright
 \addtabletext{\small{Notes: In both datasets, the first 8 central factors contain a greater amount of information $h(\emph{\textbf{X}})$, but also exhibit higher levels of noise $h(\emph{\textbf{X}}|Y)$. On the other hand, the last 8 central factors exhibit more mutual information and a higher information/noise ratio with respect to the forecast asset.}}
\end{table}

   Mutual information and information/noise ratio serve to evaluate the effectiveness of factors $\emph{\textbf{X}}$ in forecasting $Y$. They are subject to daily changes as the factors do. Table \ref{MI IN} presents the differential entropy of factors, the mutual information between the first/last 8 central factors and the forecast asset, conditional differential entropy, and information/noise ratio for each testing set, all in average sense. In both datasets, the last 8 central factors have a higher average information/noise ratio and average mutual information than the first 8 central factors for the London gold spot price. It suggests that factors with low centrality have less noise and more useful information than those with high centrality. It helps to explain why the last central factors are more accurate asset price predictors.

\begin{figure}[ht]
    \centering
    \begin{subfigure}{0.5\textwidth}
        \centering
        \includegraphics[width=\linewidth]{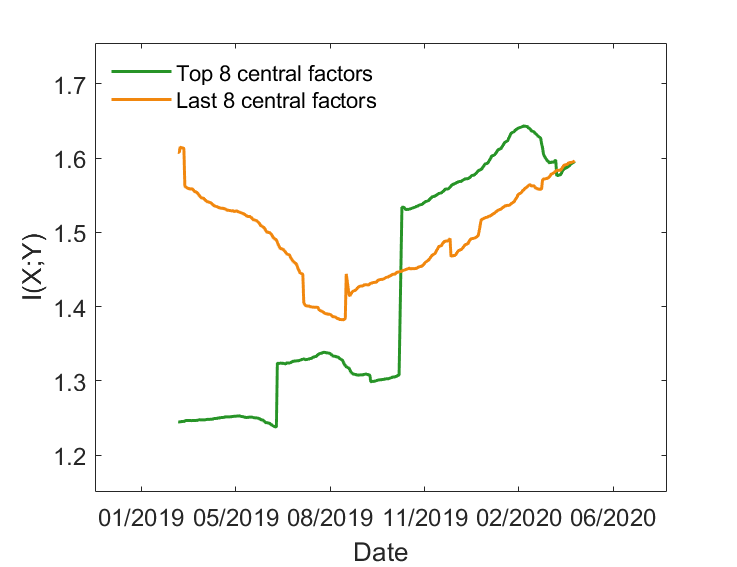}
      \subcaption{Daily mutual information.}

    \end{subfigure}%
    \hfill
    \begin{subfigure}{0.5\textwidth}
        \centering
        \includegraphics[width=\linewidth]{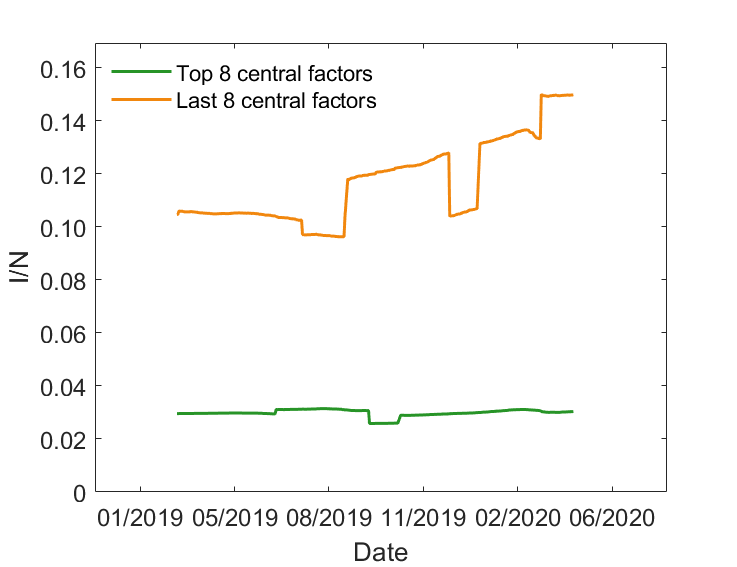}
     \subcaption{Daily information/noise ratio.}
    \end{subfigure}
    \caption{Daily mutual information and information/noise ratio in the dataset I.}
\label{MI IN fig8}
\end{figure}

Taking the dataset I as an example, the average mutual information between the last 8 central factors and the London gold spot price is greater than that between the first 8 central factors and the gold spot price during these 300 days. However, as depicted in Fig. \ref{MI IN fig8}(a), the daily mutual information between the last 8 central factors and the London gold spot price is greater than that between the first 8 central factors and the London gold spot price only on 169 out of 300 days. Observing from Fig. \ref{MI IN fig8}(b), the information/noise ratio of the last 8 central factors consistently surpasses that of the first 8 central factors. This underscores that the information/noise ratio serves as a more favorable criterion for selecting factors in forecasting scenarios. Furthermore, significant changes in Figure \ref{MI IN fig8} are attributed to the dynamic substitution of certain factors. Relatively minor changes in $I(\emph{\textbf{X}}; Y)$ and $I/N$ primarily stem from price fluctuations.

As far as now, we have explained the first problem. In general, factors with low centrality exhibit higher information/noise ratios. High-centrality factors carry more both information and noise, resulting in lower information/noise ratios and consequently inferior predictive capabilities.

\section{The impact of forecast node's centrality on prediction}
	  \label{Sec 4}

\begin{figure}[htbp]
\centering
\includegraphics[width=14cm]{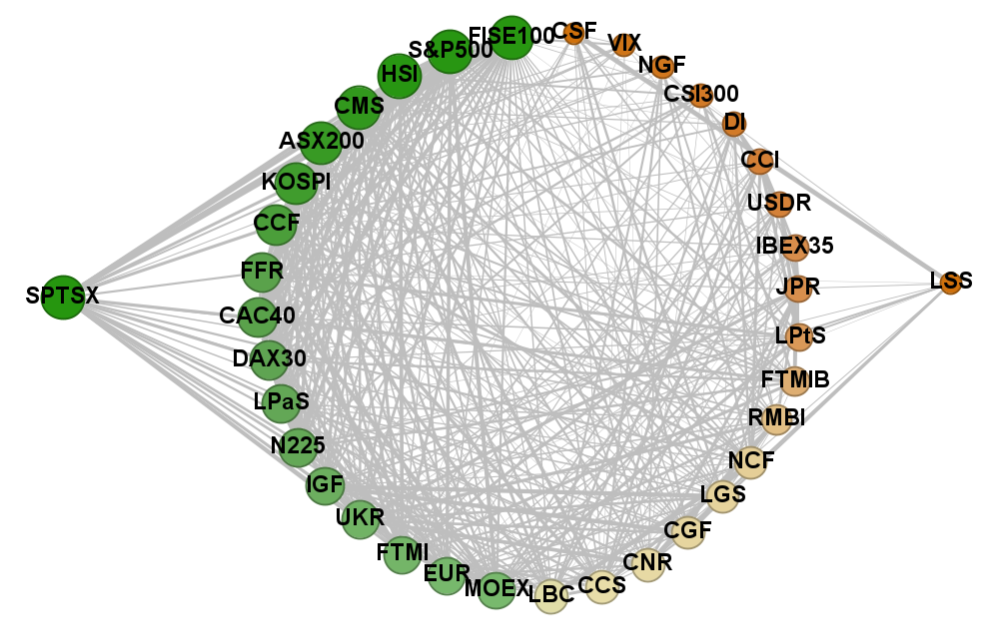}
\centering
\caption{Network on April 4, 2019. Nodes are labeled by abbreviations of the associated assets. The SPTSX on the far left has the highest eigenvector centrality. The London silver spot on the far right has the lowest eigenvector centrality. To show the relationship between nodes more clearly, we have hidden the edges with weights below 0.3. }
		\label{systemically important}
\end{figure}

In this section, we study the second problem: the impact of centrality of the predicted asset on the forecasting results.
Fig. \ref{systemically important} presents the complex network on April 4, 2019. We use the nodes on the circle to predict the price of one of the most important nodes and one of the least important nodes. Subsequently, we compare the outcomes of these predictions.

In order to highlight the influence of node centrality, in the dataset I, we selected six assets from both the top five and bottom five important nodes. In the dataset II, only four assets appeared in either the top five or bottom five important nodes more than 200 times, so we chose four assets for comparison.
Table \ref{Numbers} displays the MSINs, the LSINs, and their occurrences during 300 days.  When forecasting an asset price, we take remaining assets as factors. Due to their abundance, we extract the top 10 principal components as the actual factors instead. 


\begin{table}[htbp]	
 	\centering
 \renewcommand\arraystretch{1.2}
		\caption{The MSINs and LSINs in the two datasets.}
  		\begin{tabular}{lllll}
    \toprule
		Testing set & &\textbf{MSINs} & &\textbf{LSINs} \\
\hline
	\multirow{3}{*}{Dataset I}
	& &SPTSX (\textsl{300})  & &London silver spot(\textsl{300}) \\
 & &FISE100 (\textsl{300}) & &COMEX silver futures (\textsl{300}) \\
&   &S\&P500 (\textsl{293})& &VIX (\textsl{283})\\
     	\hline
\multirow{2}{*}{Dataset II} 	& &S\&P500 (\textsl{243})& &VIX (\textsl{241})\\ & &N225 (\textsl{233}) & &USDR (\textsl{215})\\

   \bottomrule
		\end{tabular}

  \label{Numbers}
\vspace{2mm}
\raggedright
    \addtabletext{\small{Notes: Numbers in parentheses behind assets represent the times of which these assets belong to the first/last 5 central nodes in the datasets.}}
	\end{table}

\subsection{Forecasting results for different central nodes}

 For the dataset I, Table \ref{the most important and unimportant node1} shows the forecasting results of three MSINs and three LSINs.

		 \begin{table}[htbp]
\renewcommand\arraystretch{1.2}
	\centering
      \caption{Forecasting results of different central assets in the dataset I obtained by TVF-EMD-RF.}
        \centering
		\begin{tabular}{llcccc}
  \toprule
		Centrality &Forecast asset &Dstat (\%)  &MAPE (\%) &R$^2$ &CV \\
		\midrule
      \multirow{3}{*}{\textbf{MSINs}} 	  & SPTSX& 54.33	&0.91	&0.9737  & 0.07	\\
		 &FISE100 &54.67	&1.15	&0.9811 &0.08\\
     &S\&P500 &54.67	&1.26&0.9636 &0.07   \\
  			\hline
		 \multirow{3}{*}{\textbf{LSINs}} & LSS & 59.67	&1.27	&0.9704 &0.09\\
			&CSF 	&60.33	&1.53	&0.9650  & 0.09\\
   &VIX 	&60.33	&6.72	&0.8153  &0.68  \\

			\bottomrule
		\end{tabular}

\vspace{2mm}
\justifying
      \addtabletext{\small{Notes: LSS and CSF stand for London silver spot and COMEX silver futures. We use MAPE and R$^2$ as the evaluation criteria for level accuracy because MAE and RMSE are not comparable when \\ forecasting various assets. The coefficient of variation (CV) is used to measure the degree of disp-\\ersion of their prices, which is defined as the ratio of the standard deviation to the mean.}}
		\label{the most important and unimportant node1}
	\end{table}
    The directional accuracies of the LSINs are approximately 5\%-6\% higher than those of the MSINs. However, the three LSINs are accompanied by a greater MAPE and lower R$^2$ relative to the three MSINs. The poor performance in terms of level accuracy of the three LSINs may be attributed to their large CVs. It is a well-known challenge to obtain a good level accuracy when predicting asset prices with high volatility, see \cite{LuXuZhang} for references.

     Next, we conduct an experiment to forecast two geometric Brownian movements with zero drift but varying volatilities to explain the opposite directional and level accuracy outcomes. Future values are predicted only through historical paths. The directional accuracy is roughly 50\% but level accuracy differs greatly. The geometric Brownian motion with lower volatility has a higher level accuracy. We then proceed to forecast several sets of geometric Brownian motions with varying drifts but the same volatility. The experiment result shows that a larger drift implies a higher directional accuracy, while the level accuracy is almost unchanged. This means that drift terms can be thought of as information to some extent. Hence, the directional accuracy is mainly influenced by the factors' information, whereas level accuracy is determined by both the forecast asset's volatility and the factors' information.

    These observations are also supported by our empirical analysis of the dataset II, as reported in Table \ref{the most important and unimportant  node2}. We predict two MSINs and two LSINs, respectively. The LSINs exhibit superior predictive performance compared to the MSINs in terms of directional statistics.

 \begin{table}[htbp]
 \renewcommand\arraystretch{1.2}
		\centering
		\caption{Forecasting results of different central assets in the dataset II obtained by TVF-EMD-RF.}
   \begin{tabular}{llcccc}
   \toprule
	Centrality &Forecast asset &Dstat (\%) &MAPE (\%)  &R$^2$	&CV \\
		\hline
 \multirow{2}{*}{\textbf{MSINs}} 	  & S\&P500&   53.00	&1.08	&0.9711   &0.06\\
		 &N225 &56.33	&1.16	&0.9205   &0.04 \\
   \hline
  \multirow{2}{*}{\textbf{LSINs}} &VIX &60.67	&7.93	&0.8384
  &0.22\\
   & USDR &60.67	&0.24	&0.9737  &0.01 \\
			\bottomrule
		\end{tabular}
		\label{the most important and unimportant  node2}
	\end{table}

   \subsection{Cause analysis from information perspective}

In this subsection, we make an effort to interpret and analyze the results presented above from the perspective of information rates.

Mutual information is calculated between factors (specifically, the top 10 principal components) and the assets being forecasted. Nonetheless, as highlighted in Section 3, mutual information might not be the most suitable criterion for factor selection. Given the difference of the forecasted assets, employing a metric that gauges relative information appears to be a more appropriate approach. Hence, we introduce the concept of the \emph{information rate} as the ratio of mutual information to differential entropy, defined as follows:

    	\begin{equation}
   IR=\frac{I(\emph{\textbf{X}};Y)}{h(Y)}\times 100\%=\frac{I(X_1, X_2, \cdots, X_n; Y)}{h(Y)}\times 100\%.
		\label{IR}\\
    	\end{equation}
Intuitively, $IR$ reflects the interpretable level of factors $X_1, X_2, \cdots, X_n$ to the forecast asset $Y$. It implies the proportion of useful information available in the selected factors relative to the forecast asset.

	\begin{table}[htbp]
 \renewcommand\arraystretch{1.2}
	\centering
	\caption{Average mutual information, average differential entropy, and average information rate.}
	\begin{tabular}{lllccc}
 \toprule
			\multirow{2}{*}{Testing set} &\multirow{2}{*}{Centrality}
 &\multirow{2}{*}{Asset ($Y$)}&\multicolumn{3}{c}{Daily average} \\
	\cline{4-6}
 && &$I(\emph{\textbf{X}};Y)$&$h(Y)$ &$IR$\\
		\hline
		\multirow{6}{*}{Dataset I}	&\multirow{3}{*}{\textbf{MSINs}}  &SPTSX   &2.16   &8.37  	&25.81\% \\
		&	&FISE100 &2.04   &7.59  	&26.89\% \\
      &			&S\&P500 &2.32   &9.52  	&24.32\% \\
	    	\cline{2-6}
		&\multirow{3}{*}{\textbf{LSINs}} &LSS &1.62   &1.74  	&92.83\% \\
		& &CSF &1.61   &1.75  	&91.76\% \\
  	& &VIX &1.05   &2.83  	&37.19\% \\
	\hline \hline
	\multirow{4}{*}{Dataset II} &\multirow{2}{*}{\textbf{MSINs}}&S\&P500  & 2.76  &7.63  	&36.16\% \\
 & &N225 & 2.11 &9.29  	&22.77\% \\
		\cline{2-6}
&\multirow{2}{*}{\textbf{LSINs}}  &USDR  &1.80   &2.72  &66.30\% \\
&	&VIX  &1.34  &3.40  &39.42\% \\
	\bottomrule
\end{tabular}

\vspace{2mm}
\raggedright
\addtabletext{\small{Notes: LSS and CSF stand for London silver spot and COMEX silver futures, respectively.}}
\label{Information5}
\end{table}

   Table \ref{Information5} presents the average mutual information, average differential entropy, and average information rate for the forecast assets in the 300-day testing samples across the two datasets. The most systemically important assets have more mutual information with factors; moreover, their entropy is much greater than that of the least systemically important assets. It shows that the MSINs themselves contain more information. However, the $IR$ of the MSINs is lower than that of the LSINs. This phenomenon is more pronounced in the dataset I than in the dataset II.

\begin{figure}[htbp]
		\centering
\includegraphics[width=4 in]{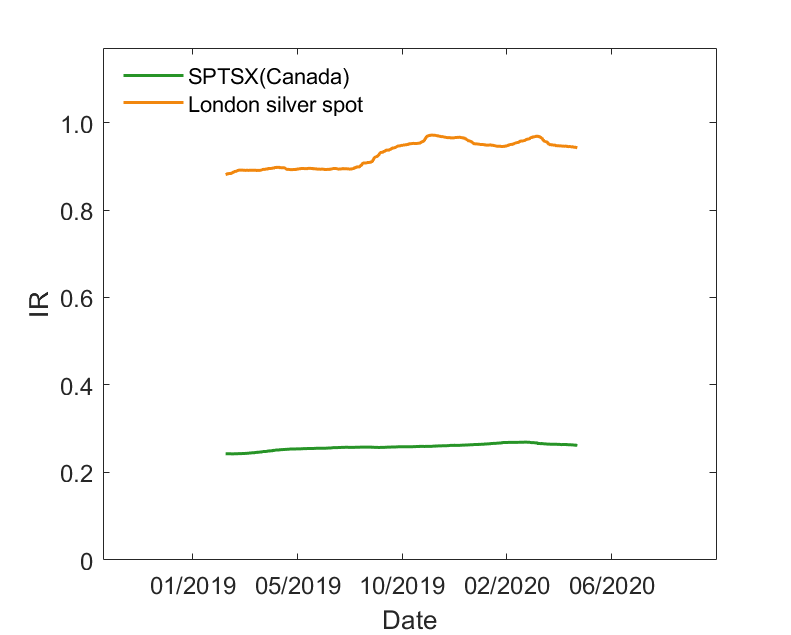}
		\centering
	\caption{Dynamic information rates of the SPTSX (the MSIN) and London silver spot (the LSIN). }
	       \label{IR fig}
	\end{figure}

    Fig. \ref{IR fig} illustrates the dynamic information rates of the SPTSX and LSS in the testing set I. The information rate of the LSS is always higher than that of the SPTSX, which is in line with the better outcomes on prediction of the LSS.

  Therefore, we can conclude that the directional accuracy of the LSINs tends to be more desirable than those of the MSINs in a closed complex system because the remaining assets have higher information rates relative to the LSINs.

\section{Robustness analysis  for the forecasting method}
  \label{Sec 5}

 In Sections 3 and 4, we have used different datasets to demonstrate the robustness of conclusions on time scales. To eliminate the contingency of the method, we use a deep extreme learning machine based on time-varying filtering-based empirical mode decomposition (TVF-EMD-DELM) to forecast asset prices. The forecasting procedures are the same with the previous sections.

For the first problem, Table \ref{I Results DELM} shows the forecasting results for the London gold spot price using different central factors in the two datasets. As we see, a set of factors with low centrality outperforms those with high centrality in both directional accuracy and level accuracy.

 \begin{table*}[htbp]
   \renewcommand\arraystretch{1.2}
		\centering
		\caption{Forecasting results of the London gold spot price obtained by TVF-EMD-DELM.}
		\begin{tabular}{llcccc}
       \toprule	
			Testing set  &Factors &Dstat (\%)&MAPE (\%) &MAE&RMSE \\
		\hline
			\multirow{6}{*}{Dataset I}	
            &First 6 central factors  &54.67 &0.91	&13.72 	&19.16  \\
			&\textbf{Last 6 central factors} &56.00 &0.94  &14.01  &23.12  \\
			\cline{2-6}
			&First 8  central factors  &54.67	&0.87 &13.07  &18.23  \\
			&\textbf{Last 8 central factors}  &58.67 &0.88	&13.26 	&19.02   \\
			\cline{2-6}
			&First 10  central factors    & 57.33	&0.93	&14.01&20.04   \\
			&\textbf{Last 10  central factors}  &60.33 &0.98	&14.74 	&20.92   \\
        \hline \hline
        \multirow{6}{*}{Dataset II}
	        &First 6  central factors   &   54.00	&0.86	&15.55	&22.82  	\\
	        &\textbf{Last 6  central factors}   &56.00	&0.84	&14.22	&19.82 \\
			\cline{2-6}
	        &First 8 central factors  & 54.67	&1.07 &19.49	&27.34 \\
 	        &\textbf{Last 8  central factors} &59.00 &1.05	&18.98 	&25.36  \\
			\cline{2-6}
	       &First 10 central factors &55.67  &0.99	&17.89 &24.41  \\
	       &\textbf{Last 10 central factors}  &57.67 &0.91	&15.58  &20.93  \\
			\bottomrule
		\end{tabular}
		\label{I Results DELM}
	\end{table*}

Table \ref{100day DELM} shows the forecasting results in twelve 100-day testing sets. In most sets, a set of factors with low centrality significantly outperforms those with high centrality in directional accuracy. Furthermore,
when the statistical window length is 200 days and 300 days, the predictive performance of nodes with low centrality is consistently better than that of nodes with high centrality.


\begin{table*}[htbp]
   \renewcommand\arraystretch{1.2}
		\centering
		\caption{Direction statistic (\%) of the London gold spot price in 12 testing sets obtained by TVF-EMD-DELM using the statistic window $w=100, 200,$ and $300$.}
		\begin{tabular}{ccccccc}
       \toprule
			\multirow{2}{*}{Testing set}   &\multicolumn{3}{c}{\textbf{First 8 central factors}} &\multicolumn{3}{c}{\textbf{Last 8 central factors}} \\
   		\cline{2-7}
 &  $w=100$ &$w=200$ & $w=300$   &$w=100$ &$w=200$ & $w=300$\\
		\hline
    1     & 57.00 & \multirow{2}{*}{57.00$^*$} & \multirow{3}{*}{56.00$^*$$^*$} & 63.00 & \multirow{2}{*}{59.50$^*$} & \multirow{3}{*}{58.33$^*$$^*$} \\
    2     & 57.00 &       &       & 56.00 &       &  \\
    3     & 54.00 & \multirow{2}{*}{54.50$^*$} &       & 56.00 & \multirow{2}{*}{58.50$^*$} &  \\
    4     & 55.00 &       & \multirow{3}{*}{53.67$^*$$^*$} & 61.00 &       & \multirow{3}{*}{59.33$^*$$^*$} \\
    5     & 52.00 & \multirow{2}{*}{53.00$^*$} &       & 58.00 & \multirow{2}{*}{58.50$^*$} &  \\
    6     & 54.00 &       &       & 59.00 &       &  \\
    7     & 58.00 & \multirow{2}{*}{54.00$^*$} & \multirow{3}{*}{53.33$^*$$^*$} & 59.00 & \multirow{2}{*}{59.00$^*$} & \multirow{3}{*}{58.00$^*$$^*$} \\
    8     & 50.00 &       &       & 59.00 &       &  \\
    9     & 52.00 & \multirow{2}{*}{52.00$^*$} &       & 56.00 & \multirow{2}{*}{57.00$^*$} &  \\
    10    & 52.00 &       & \multirow{3}{*}{54.67$^*$$^*$} & 58.00 &       & \multirow{3}{*}{59.00$^*$$^*$} \\
    11    & 57.00 & \multirow{2}{*}{56.00$^*$} &       & 60.00 & \multirow{2}{*}{59.50$^*$} &  \\
    12    & 55.00 &       &       & 59.00 &       &  \\
     \bottomrule
     \end{tabular}

     \addtabletext{\small{Notes:$^*$ The statistic window length is $w=200$; $^*$$^*$ The statistic window length is $w=300$.}}
		\label{100day DELM}
	\end{table*}

For the second problem, Table \ref{II DELM} displays the forecasting results of different central assets in the two datasets using TVF-EMD-DELM. Directional accuracies of the LSINs are 4\%-8\% higher than that of the MSINs. All these conclusions are consistent with those obtained by TVF-EMD-RF.

 \begin{table}[htbp]
 \renewcommand\arraystretch{1.2}
		\centering
		\caption{Forecasting results of different central assets obtained by TVF-EMD-DELM.}
		\begin{tabular}{lllccc}
  \toprule
	Testing set &Centrality &Forecast asset &Dstat (\%) &MAPE (\%)   &$R^2$	
   \\
				\hline
 \multirow{6}{*}{\small{Dataset I}}      &\multirow{4}{*}{\textbf{MSINs}} & \small{SPTSX}   &54.00\%	&1.21\%	&0.9066     \\
		& 	&\small{FISE100}  &  56.00\%	&1.21\%	&0.8975    \\
&	&\small{S\&P500} &55.00\%	&1.24\%	&0.9378 	\\
  			\cline{2-6}
	&	\multirow{4}{*}{\textbf{LSINs}} & \small{London silver spot}&59.33\%	&1.66\%	&0.9014
\\
   	&	&\small{COMEX silver futures} &59.33\%	&1.66\%	&0.8897  \\
&	&\small{VIX} &60.00\%	&7.56\%	&0.8388 \\
   \hline \hline
	\multirow{4}{*}{\small{Dataset II}}       	& \multirow{2}{*}{\textbf{MSINs}} &\small{S\&P500}  &54.33\%	&1.01\%	&0.9103    \\
&&\small{N225} &54.00\%	&1.34\%	&0.8439 \\
         \cline{2-6}
	    &  \multirow{2}{*}{\textbf{LSINs}}	 & \small{VIX}      & 59.00\%	&9.04\%	&0.7139       \\
  &   & \small{USDR}     & 59.33\%	&0.21\%	&0.9189  \\
			\bottomrule
		\end{tabular}
		\label{II DELM}
	\end{table}

As now, we have verify our results using different datasets, two methods of machine learning, and various nodes.

\section{Conclusion}
  \label{Sec 6}

This paper studies the impacts of node centrality on asset price prediction in networks of major global assets from two aspects: factors' centrality and the forecast node's centrality. There are usually two intriguing conclusions. First, a set of factors with low centrality typically produces more satisfying forecasting outcomes than a set of factors with high centrality when forecasting the London gold spot. Second, the LSINs tend to be predicted more accurately than the MSINs in terms of directional accuracy. We verify the robustness of these conclusions using different datasets and different machine learning methods.

	We explain these phenomena from the point of view of information. On the one hand, we find that the set of factors with low centrality usually provides more effective information for the forecast asset employing information theory. Also, it is discovered that the set of factors with low centrality has a high information/noise ratio. Furthermore, the forecasting results depend more on the information/noise ratio than mutual information. On the other hand, we demonstrate that the directional accuracy of various forecast assets rests with the information rate, even though factors have more mutual information with the MSINs than with the LSINs.

    In general, nodes with high centrality are important in a complex network. For instance, the risk of a bank system is almost determined by systemically important banks. However, our conclusions suggest that forecasting results improve with decreasing centrality of factors and forecast nodes. This sheds light on factor selection by showing that nodes with low centrality in a complex network, rather than only SINs, should be considered when predicting a node. It's worth noting that our conclusions are more significant only when there is a significant difference in nodes centrality. Less distinct differences in nodes centrality can weaken our conclusions.

    One direction for future work is to expand the application of our findings to networks consisting of non-financial assets to obtain more general conclusions. Another possibility is to consider various relationships among nodes and alternative measures of centrality.





\appendix

\section{Data}\label{appendix:Data}

Table \ref{data source} lists the names, categories, abbreviations, and sources of the data for assets in the complex network.

\setcounter{table}{0} 

\begin{table*}[htbp]
\renewcommand\arraystretch{1.2}
		\centering
		\caption{Data categories and sources.}
    {\fontsize{9}{12}\selectfont
  \begin{tabular}{lllll}
   	\toprule
				Number & Category &Name &Abbreviation &Source \\
				\hline 
				1&Commodity & London gold spot &LGS& https://www.wind.com.cn\\
				2&Commodity &London silver spot&LSS & https://www.wind.com.cn\\
				3&Commodity &London palladium spot&LPdS&
                              https://www.kitco.com/gold.londonfix.html\\
				4&Commodity &London platinum spot &LPtS&
                              https://www.kitco.com/gold.londonfix.html\\
			    5&Commodity &COMEX gold futures&CGF& https://www.wind.com.cn\\
				6&Commodity &COMEX silver futures &CSF& https://www.wind.com.cn\\
				7&Commodity &COMEX copper futures &CCF& https://www.wind.com.cn\\
				8&Commodity &ICE gold futures&IGF& https://www.wind.com.cn\\
				9&Commodity &CRB metal spot &CMS& https://www.wind.com.cn\\
				10&Commodity &CRB commodity index&CCI & https://www.wind.com.cn\\
				11&Commodity &CRB comprehensive spot&CCS& https://www.wind.com.cn\\
				12&Commodity &NYMEX gas futures &NGF & https://www.wind.com.cn\\
				13&Commodity &NYMEX crude futures&NCF& https://www.wind.com.cn\\
				14&Commodity &London brent crude&LBC& https://www.wind.com.cn\\
				15&Stock index  & ASX200 (Australia)&ASX200& https://www.wind.com.cn\\
				16&Stock index  & SPTSX (Canada)& SPTSX & https://www.wind.com.cn\\
				17&Stock index & CSI300 (China)&CSI300 & https://www.wind.com.cn\\
				18&Stock index  & MOEX (Russia)&MOEX& https://www.wind.com.cn\\
				19&Stock index  & CAC40 (France)& CAC40 & https://www.wind.com.cn\\
				20&Stock index  & HSI (Hong Kong)&HSI& https://www.wind.com.cn\\
				21&Stock index  & IBEX35 (Spain)&IBEX35 & https://www.wind.com.cn\\
				22&Stock index  & FTMIB (Italy)&FTMIB& https://www.wind.com.cn\\
				23&Stock index  & KOSPI (Korea)&KOSPI& https://www.wind.com.cn\\
				24&Stock index  & FTMI (Singapore) &FTMI & https://www.wind.com.cn\\
				25&Stock index  & FISE100 (UK) &FISE100& https://www.wind.com.cn\\
				26&Stock index  & S\&P500 (US)& S\&P500 & https://www.wind.com.cn\\
				27&Stock index  & N225 (Japan)&N225& https://www.wind.com.cn\\
				28&Stock index  & DAX30 (Germany) & DAX30 & https://www.wind.com.cn\\
				29&Economic indicator &VIX &VIX& https://www.wind.com.cn\\
				30&Economic indicator &Dollar index &DI& https://www.wind.com.cn\\
				31&Economic indicator &RMB index&RMBI& https://www.wind.com.cn\\
				32&Economic indicator &UK effective exchange rate   &UKR &https://www.bis.org/statistics/eer.htm      \\
				33&Economic indicator &USD effective exchange rate     &USDR&https://www.bis.org/statistics/eer.htm      \\
				34&Economic indicator &EU effective exchange rate   &EUR &https://www.bis.org/statistics/eer.htm      \\
				35&Economic indicator &CN effective exchange rate      &CNR&https://www.bis.org/statistics/eer.htm      \\
				36&Economic indicator &JP effective exchange rate   &JPR&https://www.bis.org/statistics/eer.htm      \\
				37&Economic indicator &Federal funds effective rate &FFR &https://www.federalreserve.gov	 \\	
				\bottomrule
    			\end{tabular}
}
  \label{data source}
	\end{table*}

\section{Some methods}\label{appendix:Methods}

We briefly introduce some methods, including random forest, TVF-EMD, and principal component analysis.

\subsubsection*{B.1 Random forest}
    \cite{Ho95, Ho98} developed the basic idea of a random forest. As an ensemble learning model, the random forest combines Breiman's ``bagging'' idea \citep{Breiman1996} and a random selection of features. It takes the classification and regression trees (CARTs) as the basic model. CART is a binary decision tree \citep{BreimanFriedman}. For a classification problem, it uses the Gini index as a criterion to choose features and the splitting points. For a regression problem, it uses the mean square error (MSE) as a splitting criterion.

    The final results of a random forest depend on the number of votes in all decision trees. For a classification problem, the random forest takes the majority vote of all decision trees as its result. For a regression problem, it uses the mean value. The random forest has the advantages of high speed and accuracy. More details are referred to \cite{Breiman3}.

\subsubsection*{B.2 TVF-EMD}

  Empirical mode decomposition (EMD) is a classic signal processing approach. Given an original signal $s(t)$, one can apply EMD to decompose it into different frequency bands in the time domain, known as intrinsic mode functions (IMFs). However, the method has the disadvantages of small end effects and mode aliasing. \cite{LiLiMo} proposed the TVF-EMD method based on time-varying filters, which is an adaptive algorithm that does not require a manual determination of the parameters. Compared with EMD, TVF-EMD can improve the frequency separation performance and stability at low sampling rates. In addition, the proposed method is robust against noise interference. Therefore, we adopt TVF-EMD to decompose the price series in this paper.

\subsubsection*{B.3 Principal component analysis}

 Principal component analysis (PCA) is a popular technique for analyzing large datasets containing a high number of dimensions/features per observation. PCA is used to reduce the dimensionality of a dataset while preserving the maximum amount of information. It implies that the greater the variance, the greater the amount of information.

 PCA using the covariance method requires the following steps \citep{Hotelling, Pearson}. (i) Centralize all original features. (ii) Calculate covariance matrix $C$. (iii) Find the eigenvalues of $C$ and the corresponding eigenvectors. (iv) Select the first largest $k$ eigenvalues and the corresponding eigenvectors. (v) Project the original features onto the selected eigenvectors and obtain new $k$-dimensional features after reducing the dimensions.

 The percentage of the sum of the first largest $k$ eigenvalues in the sum of all eigenvalues is called the contribution rate. Usually, the top several principal components with a given contribution rate (such as 80\%) or a certain number of the first several components are adopted.

\section{Several concepts of information theory} \label{appendix:information}

\subsection*{C.1 Shannon entropy and differential entropy}
	 \cite{Shannon} proposed the concept of \textit{entropy}, where the entropy $H(X)$ of a discrete random variable $X$ with possible values ${x_1, x_2, \cdots, x_n}$ and the probability mass function $P(X)$ is defined as

	\begin{equation}
		H(X):=-\sum_{i=1}^{n}P(x_i)\log_b P(x_i).
		\label{H(X)}
	\end{equation}
	This reflects the expected amount of information of $X$. Otherwise, the entropy is also a measure of the uncertainty of a random variable. The frequently-used values of $b$ are 2, 10, and $e$, and we use $e$ as the base in this paper.
	
    Correspondingly, \textit{differential entropy} (also called continuous entropy) describes continuous probability distributions. Let $X$ be a continuous random variable with a probability density function $p$ whose support is a set $\mathcal{X}$. Its differential entropy $h(X)$ is defined as

   	\begin{equation}
		h(X):=-\int_\mathcal{X} p(x)\ln p(x)\,dx.  \notag
	\end{equation}

    It can be used to compare the uncertainty of two continuous random variables. Compared with the entropy of discrete random variables, the differential entropy can be negative and is not invariant under continuous coordinate transformations. Therefore, we should compare the uncertainty of two continuous random variables at the same scale.

\subsection*{C.2  Conditional differential entropy}

    \textit{Conditional differential entropy} quantifies the amount of information required to describe the outcome of a continuous random variable given the value of another continuous random variable. Let $X$ and $Y$ be continuous random variables with values ranging in $\mathcal{X}$ and $\mathcal{Y}$ with a joint probability density function $p(x,y)$. Their marginal probability density functions are $p(x)$ and $p(y)$, respectively. The conditional differential entropy $h(Y|X)$ is defined as

   	\begin{equation}
		h(Y|X):=-\int_\mathcal{Y}\int_\mathcal{X} p(x,y)\ln\frac{p(x,y)}{p(x)}\,dxdy.
	\end{equation}
	
\subsection*{C.3 Mutual information}

      \textit{Mutual information} is a measure of the mutual dependence between two random variables.
   The mutual information between two continuous random variables, $X$ and $Y$, is defined as

  	\begin{equation}
   \begin{aligned}
		I(X;Y):&=h(Y)-h(Y|X)\\
          &=h(X)-h(X|Y)\\&=\int_\mathcal{Y}\int_\mathcal{X} p(x,y)\ln\frac{p(x,y)}{p(x)p(y)}\,dxdy.
	\label{MI2}
 \end{aligned}
        \end{equation}
   It is equal to the difference between the differential entropy and conditional differential entropy.

   The intuitive meaning of mutual information is the amount of information (i.e., reduction in uncertainty) that one variable provides to the other. Hence, mutual information is also known as information gain.




\end{document}